# Exploring topological phase transition and Weyl physics in five dimensions with electric circuits

Xingen Zheng ⬚,* Tian Chen ⬚,*,† Weixuan Zhang, Houjun Sun ⬚, and Xiangdong Zhang ⬚‡

*Key Laboratory of Advanced Optoelectronic Quantum Architecture and Measurements of Ministry of Education,*
*Beijing Key Laboratory of Nanophotonics & Ultrafine Optoelectronic Systems, School of Physics,*
*Beijing Institute of Technology, Beijing 100081, China*

Weyl semimetals are phases of matter with gapless electronic excitations that are protected by topology and symmetry. Their properties depend on the dimensions of the systems. It has been theoretically demonstrated that five-dimensional (5D) Weyl semimetals emerge as novel phases during the topological phase transition in analogy to the three-dimensional case. However, experimental observation of such a phenomenon remains a great challenge because the tunable 5D system is extremely hard to construct in real space. Here, we construct 5D electric circuit platforms in fully real space and experimentally observe topological phase transitions in five dimensions. Not only are Yang monopoles and linked Weyl surfaces observed experimentally, but various phase transitions in five dimensions are also proved, such as the phase transitions from a normal insulator to a Hopf link of two Weyl surfaces and then to a 5D topological insulator. The demonstrated topological phase transitions in five dimensions leverage the concept of higher-dimensional Weyl physics to control electrical signals in the engineered circuits.

## I. INTRODUCTION

The discovery of topological states of matter has greatly enriched the variety of matter in nature, and the exploration of topological physics in various systems has become one of the most fascinating frontiers in recent years [1–4]. In three dimensions, a significant topological state is the Weyl semimetal, which has chiral anomaly and topological Fermi arc surface states [5–9]. It has a key role in topological phase transitions between topological states involving a change of topology of the ground-state wave function [10–14]. Instead of being described by local order parameters, topological phases are described by global topological invariants and are robust against perturbations. Thus, the study of topological states has greatly improved our understanding of phase transitions in physical systems.

Recent investigations have been extended to higher-dimensional systems ($n > 3$), including the four-dimensional (4D) quantum Hall effect [15–19], Yang monopoles [20,21], two-dimensional (2D) Weyl surfaces in five-dimensional (5D) systems [22–27], and so on. The higher-dimensional systems are expected to possess properties that their lower-dimensional counterparts do not support. It has been theoretically verified that 5D Weyl semimetals emerge as novel phases during the topological phase transition, which carry a non-Abelian second Chern number for Yang monopoles and a linking number for linked Weyl surfaces. However, experimental observation of such a phenomenon remains a challenge, although linked Weyl surfaces and Weyl arcs can be observed in photonic metamaterials and atomic Bose-Einstein condensates [20,27]. To observe the topological phase transition containing these two intermediate phases in 5D, it is necessary not only to construct a 5D physical system but also to have parameter-adjustable characteristics. At present, no realizable cases have been proposed yet.

In this works, we construct 5D circuit platforms in fully real space and experimentally study topological phase transitions and Weyl physics in 5D. Not only are Yang monopoles and surface Weyl arcs investigated experimentally, but also the 5D Weyl semimetals emerging as intermediate phases during the topological phase transition are also observed. Our work opens up the exciting possibility of exploring Weyl physics in high-dimensional topology.

## II. 5D WEYL SEMIMETAL CIRCUITS

To construct 5D Weyl semimetal circuits, we consider structures as shown in Fig. 1(a), in which the plaquettes are marked by red dots. Each plaquette contains four circuit nodes (A–D). The advantage of the circuit is that we only need to consider the connection relationship between different circuit nodes instead of their location [28–43]. Thus, the 5D semimetal structure can be constructed by properly connecting the circuit components along five directions. The solid and dashed (orange, blue, brown, green, and pink) lines represent the connection couplings among nodes along the $x$, $y$, $z$, $u$, and $v$ directions, respectively. By extending such connections with designed circuit components, the 5D Weyl semimetal circuit containing $N_x \times N_y \times N_z \times N_u \times N_v$ plaquettes can be obtained. Figure 1(a) shows such a circuit containing $2 \times 2 \times 2 \times 2 \times 2$ plaquettes; the corresponding couplings realized by inductors and capacitors between adjacent plaquettes along five directions are given in Fig. 1(b). Two kinds of inductors and capacitors are used in the circuit, the magenta (purple) inductor $L_1(L_2)$, the green (yellow) capacitor $C_1(C_2)$, and $C_2 : C_1 = L_1 : L_2 = \sqrt{2} : 1$. In Fig. 1(b), the admittance between two nodes can be changed by switching the two components in series along the $z$ direction. in addition, proper grounding elements should be adjusted according to the different contents we want to observe, which include the Yang monopole, the linked Weyl surfaces, and topological phase transition. The corresponding grounding elements for observing these three cases are shown in Figs. 1(c)–1(e), respectively.

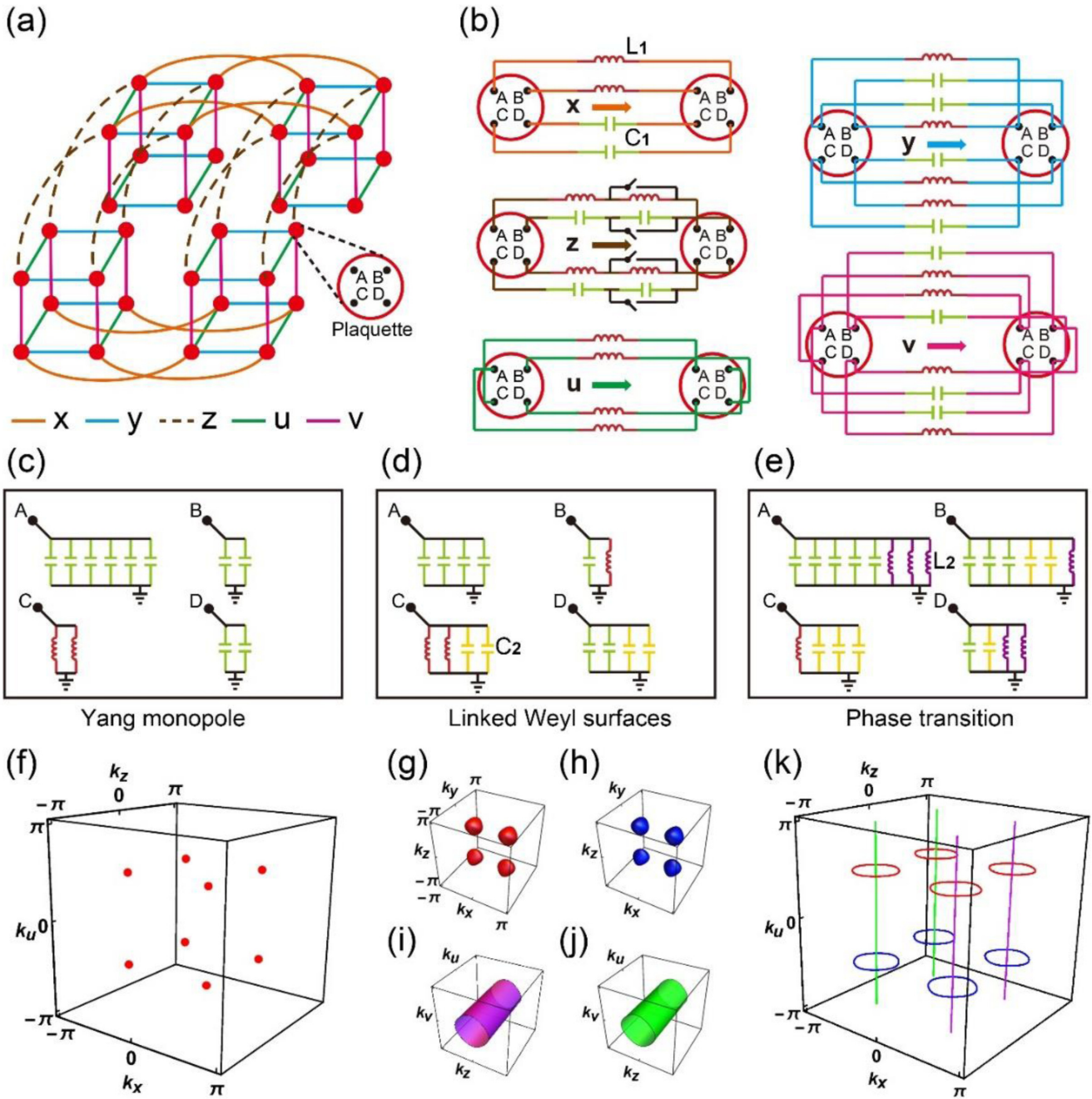

FIG. 1. (a) Connection relationship in our 5D circuit, in which the red dots represent the plaquettes, and the orange, blue, brown, green, and pink lines represent the couplings in the $x$, $y$, $z$, $u$, and $v$ directions, respectively. (b) The specific couplings realized by inductors and capacitors between adjacent plaquettes in each direction. (c)–(e) Different grounding conditions for different phases. (f) Eight Yang monopoles shown in $k_y = k_v = 0$ subspace. (g)–(j) Weyl surfaces in different subspaces. (k) The linking relationship of Weyl surfaces in $k_y = k_v = 0$ subspace.

For the above circuit network, we can derive a circuit Laplacian (or, say, admittance matrix) by Kirchhoff's law.

In a matrix form, Kirchhoff's law is expressed as $\mathbf{I} = J\mathbf{V}$, where the column vectors $\mathbf{V}$ and $\mathbf{I}$ represent the voltage and the current at each node, and $J$ is the circuit Laplacian. The circuit Laplacian $J(\mathbf{k})$ in momentum space at the resonance frequency $\omega_0 = 1/\sqrt{L_1 C_1}$ can be derived as

$$J(\mathbf{k}) = 2i\sqrt{\frac{C_1}{L_1}}\left\{\sum_{i=1}^{5}\zeta_i(\mathbf{k})\Gamma^i + ia\frac{[\Gamma^4, \Gamma^5]}{2}\right\}, \qquad (1)$$

where $\Gamma^i$ are the five $4\times4$ Gamma matrices satisfying the anticommutation relation and $0 < a < 1$. We choose the representation $\Gamma^{1,2,3,4,5} = \{\sigma_3\tau_1, \sigma_3\tau_2, \sigma_3\tau_3, \sigma_1, \sigma_2\}$, where $\sigma$ and $\tau$ are both Pauli matrices. The functions $\xi_i(\mathbf{k})$ are defined as $\zeta_1(\mathbf{k}) = \cos k_x$, $\zeta_2(\mathbf{k}) = \sin k_y$, $\zeta_3(\mathbf{k}) = \eta(\cos k_z - 1) + (\cos k_y + \cos k_v - 2) + m$, $\zeta_4(\mathbf{k}) = \cos k_u$, and $\zeta_5(\mathbf{k}) = \sin k_v$, where $\eta$ and $m$ are adjustable real parameters. These adjustable parameters can be determined by using the switches and different grounding conditions in circuits. The detailed derivation of Eq. (1) is provided in Appendix A. Then, the Hamiltonian of the circuit system can be obtained as $H(\mathbf{k}) = J(\mathbf{k})/(2i\sqrt{C_1/L_1})$, which corresponds exactly to the form of the 5D Weyl semimetal lattice. The energy spectrum of $H(\mathbf{k})$ can be easily obtained as $\varepsilon_i = \pm\{[(\zeta_1^2 + \zeta_2^2 + \zeta_3^2)^{1/2} \pm a]^2 + \zeta_4^2 + \zeta_5^2\}^{1/2}$, where $\varepsilon_i$ ($i = 1, 2, 3, 4$) stands for the $i$th lowest band.

Based on the circuit above, we can observe a series of phenomena for 5D Weyl physics in which Yang monopole and linked Weyl surfaces are two of the most important topological phases. To illustrate these two phases, we take $\eta = 1$ and $m = 1$, then $\zeta_3(\mathbf{k}) = \cos k_z + (\cos k_y + \cos k_v - 2)$. When $a = 0$ is chosen, the four bands form eight fourfold degenerate points in momentum space, locating at $k = (\pm\frac{\pi}{2}, 0, \pm\frac{\pi}{2}, \pm\frac{\pi}{2}, 0)$ as shown in the subspace of $k_y = k_u = 0$ in Fig. 1(f). At these eight degenerate points, the corresponding Hamiltonian can be expanded into the standard form of the 5D Dirac equation, which means these points are Yang monopoles. The Yang monopole can be regarded as the monopole of Berry curvature in the 5D space with a nonzero second Chern number $C_2$ [23]. The calculation method of the second Chern number is described in Appendix B.

When the perturbation is introduced (in our case, we set $a = 1/\sqrt{2}$), the time reversal combined with space inversion symmetry is broken, and the fourfold degenerate points in momentum space disappear. In this case, the intersection of band $\varepsilon_2$ and band $\varepsilon_3$ forms eight spherelike surfaces which can be seen in the subspace of $k_u = \pi/2(-\pi/2)$ and $k_v = 0$, as shown in Figs. 1(g) and 1(h). At the same time, the intersection of band $\varepsilon_3$ and band $\varepsilon_4$ (or band $\varepsilon_1$ and band $\varepsilon_2$) forms two tori in the subspace of $k_u = \pi/2(-\pi/2)$ and $k_y = 0$, as shown in Figs. 1(i) and 1(j). These surfaces are Weyl surfaces since they are twofold degenerate and can be expanded into the Weyl equation. Also there exists linking relationships between these Weyl surfaces which can be seen in the subspace of $k_y = k_u = 0$ as shown in Fig. 1(k). The red, blue, purple, green lines in Fig. 1(k) correspond to the Weyl surfaces with the same color in Figs. 1(g)–1(j), respectively. It has been proved that the second Chern number of the Weyl surface is equal to the linking number in such a case [22].

Therefore, we can calculate the second Chern number as the topological invariant (see Appendix B for detail).

## III. OBSERVATION OF TOPOLOGICAL PHASES IN 5D WEYL SEMIMETAL CIRCUITS

To observe the above phenomena experimentally, the designed electric circuit with $8\times2\times8\times4\times2$ plaquettes is fabricated. The description of the design principle and fabrication method is given in Appendix C. Based on the fabricated circuit network, the admittance spectrum of the system in momentum space can be obtained experimentally by an elementary voltage measurement and the Fourier transform. The detailed measurement method is described in Appendix D. When the grounding condition is chosen according to the way shown in Fig. 1(c), and the switches are turned off to ensure $\eta = 1$, the measured results are described by red dots in Fig. 2(a).

Figure 2(a) shows the spectrum in the 2D subspace formed by $k_x$ and $k_z$ with $k_y = k_v = 0$ and $k_u = \pi/2$. The translucent surfaces in Fig. 2(a) display the corresponding theoretical results. In this paper, all the theoretical results are calculated with $64\times64\times64\times64\times64$ plaquettes. There is only one coefficient difference between the circuit Laplacian and the Hamiltonian; therefore, the translucent surfaces can also be regarded as the theoretical spectrum of the tight-binding model. Comparing the red dots and the translucent surfaces in Fig. 2(a), we find that the agreement between theoretical and experimental results is very good. This can be seen more clearly from Fig. 2(b). The four panels in Fig. 2(b) correspond to the sectional views of four different $k_x$ values for Fig. 2(a), respectively. Solid lines (theory) and red dots (experiment) show good consistency even though some errors exist. These errors come from the inaccuracy of the inductors and capacitors used and some effects of the long wires. It is also seen that there are four degenerate points (the first and third panels) in such a case with $k_u = \pi/2$. For the case with $k_u = -\pi/2$, a similar spectrum with the other four degenerate points can be observed. These eight degenerate points correspond to the eight red dots in Fig. 1(f); that is, Yang monopoles are demonstrated experimentally.

While in the 5D system with Yang monopoles, when the boundary of the lattice is open along an appropriate direction, here we choose the $v$ direction, projecting the Yang monopoles into a 4D surface; an arc connecting a pair of Yang monopoles also appears, which is called a Weyl arc [23]. Accompanied by the Weyl arc is the boundary state on the 4D surface. In order to observe such a phenomenon experimentally, we remove the electrical elements between the head and tail in the $v$ direction and adjust the grounding elements at the head and tail accordingly to keep the diagonal terms of the circuit Laplacian $J$ unaffected. For such an open boundary system, the calculated spectrum of the circuit Laplacian as a function of driving frequency is shown in Fig. 2(c). It is clearly seen that the zero mode, which corresponds to the boundary state, appears in the spectrum gap near the resonant frequency $\omega_0$. In circuit, the zero mode usually manifests an impedance peak [31], and it is also observed in our experiment. As shown in Fig. 2(d), the simulated results of frequency-dependent impedances are plotted by the red line; the corresponding experimental results are described by the blue line. Both theoretical and experimental results exhibit the resonant peaks around $\omega_0$ in the gap, which are identical with the spectrum shown in Fig. 2(c).

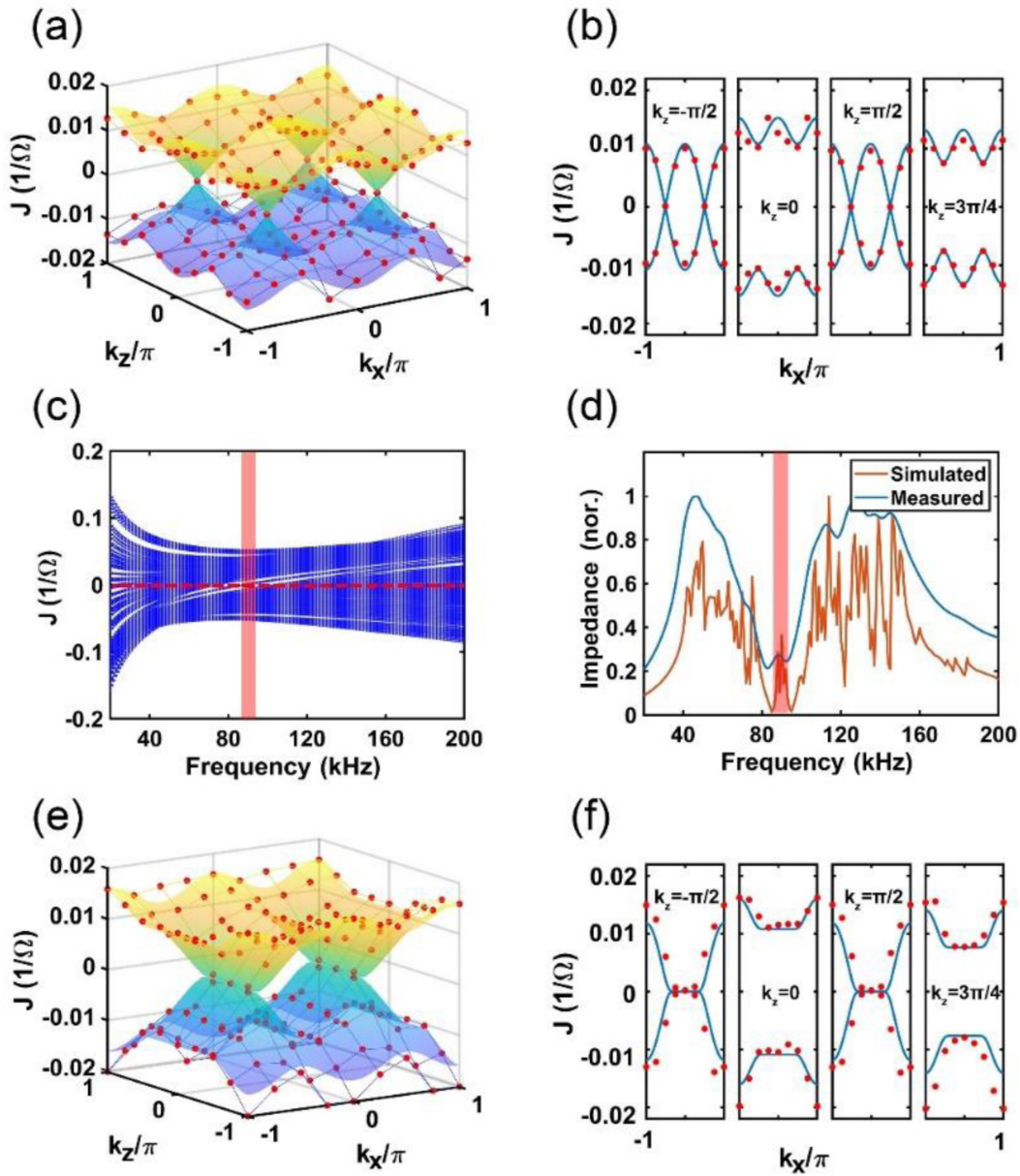

FIG. 2. (a) Measured admittance spectrum of the Yang monopole system in the subspace of $k_x$ and $k_z$. (b) Four slices of the measured admittance spectrum of the Yang monopole system in $k_x$ subspace. (c) The spectrum of the circuit Laplacian $J$ as a function of driving frequency. (d) Impedance values measured and simulated at the 4D boundary of the Yang monopole system. (e) The Weyl arcs observed after projecting the Yang monopole system into the 4D surface. (f) Four slices in $k_x$ subspace corresponding to (e).

The measured admittance spectrum of the open system is described by red dots in Fig. 2(e), where the 2D subspace is also formed by $k_x$ and $k_z$ with $k_y = k_v = 0$ and $k_u = \pi/2$. The translucent surfaces in Fig. 2(e) also represent the corresponding theoretical results. The Weyl arcs connected between the original Yang monopoles are observed. To show the phenomenon more clearly, in Fig. 2(f) we provide the sectional views of four different $k_x$ values for Fig. 2(e). The Weyl arcs are clearly displayed in the first and third panels.

The agreements between theoretical results (solid lines) and experimental measurements (red dots) are observed again.

After adjusting the grounding elements to the case in Fig. 1(d), the corresponding experiments of linked Weyl surfaces are performed, and the measured results are shown in Figs. 3(a) and 3(b) for $k_u = \pi/2$ and $k_u = -\pi/2$, respectively. Here, the 2D subspaces of $k_x$ and $k_z$ are also taken. We observe four Weyl surfaces in Fig. 3(a) (red circles) and Fig. 3(b) (blue circles), respectively, which correspond to four red spheres in Fig. 1(g) and four blue spheres in Fig. 1(h). The 2D torus (green lines) also appears, which corresponds to the cases in Figs. 1(i) and 1(j). In order to show the phenomenon better, we draw the points on the Weyl surfaces alone in Figs. 3(c) and 3(d). Figure 3(c) corresponds to the case in the 2D subspace of $k_x$ and $k_z$, and Fig. 3(d) describes the case in the 2D subspace of $k_x$ and $k_u$, in which this 2D subspace can be regarded as the projections of Fig. 1(k). This means that our experiments well prove the theoretical results. Furthermore, to observe the Weyl arc in the linked Weyl surfaces system, we still open the boundary in the $v$ direction. The theoretical (translucent surface) and experimental (red dots) results are shown in Fig. 3(e). Here, the spectrum is shown only below the Fermi surface because the spectra are symmetrical with respect to zero energy. Even the number of plaquettes along the $v$ directions is a little small; the Weyl arcs still can be seen clearly in this case. The corresponding points exactly on the Weyl surfaces in the 2D subspace of $k_x$ and $k_z$ are plotted in Fig. 3(e). The experimental deviation on Weyl arcs in Fig. 3(e) is less than 0.005 $1/\Omega$. We give a detailed discussion about the effect of the sample size in Appendix E, where more theory calculation results and simulated results for various sample sizes are provided.

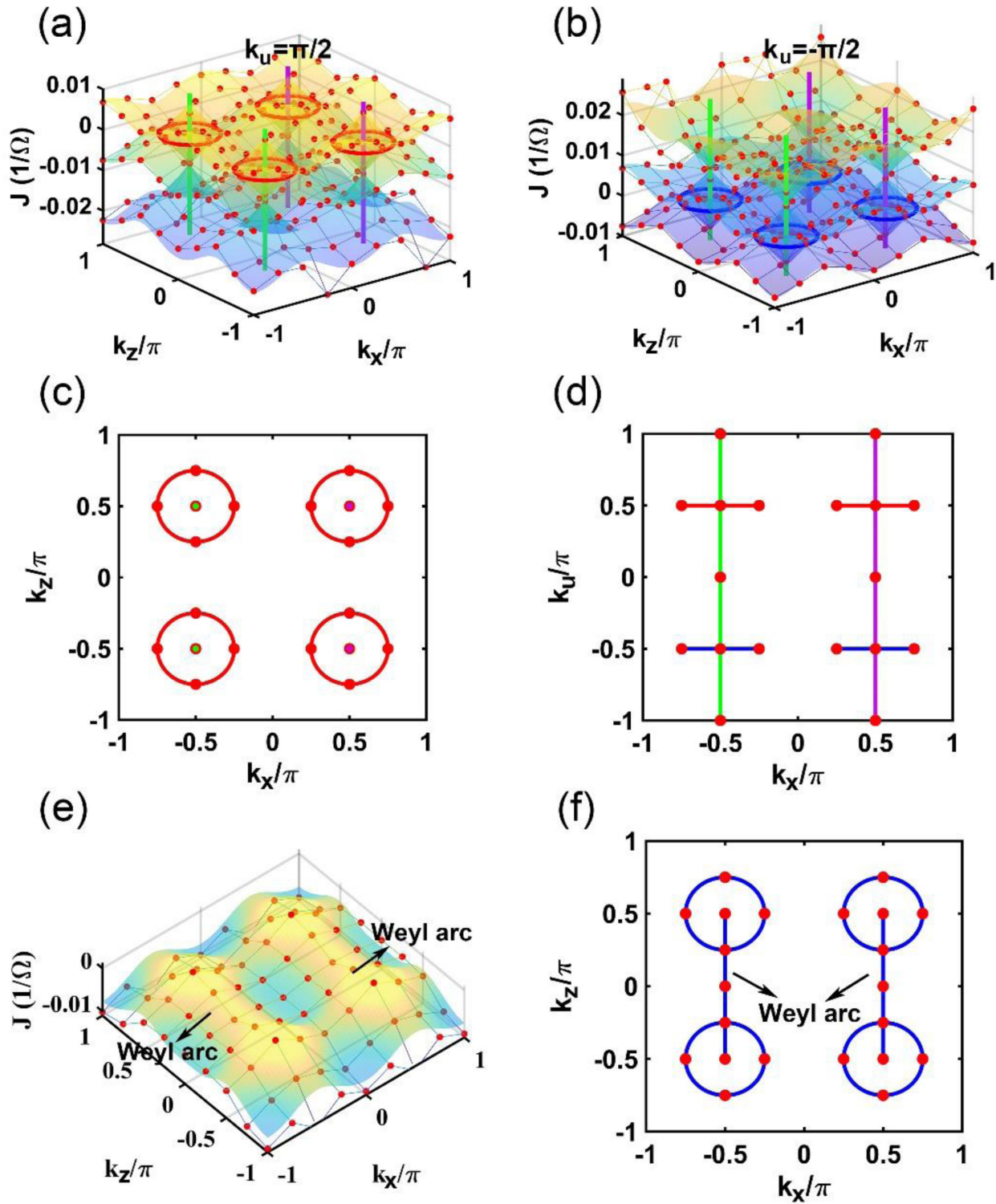

FIG. 3. (a), (b) Measured admittance spectrum of the linked Weyl surfaces system in the subspace of $k_x$ and $k_z$. (c) The dots on Weyl surfaces observed in $k_x$ and $k_z$ subspace. (d) The dots on Weyl surfaces observed in $k_x$ and $k_u$ subspace. (e) The Weyl arcs observed after projecting the linked Weyl surfaces system into the 4D surface. (f) The measured spectrum at the Fermi surface.

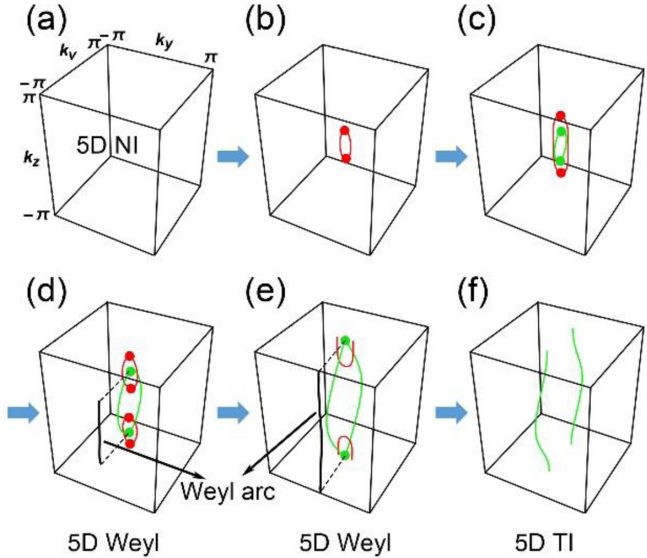

FIG. 4. This process is plotted in the 3D subspace of $k_x = k_u = \pi/2$. The red lines are Weyl surfaces between bands $\varepsilon_2$ and $\varepsilon_3$, the green lines are those between bands $\varepsilon_3$ and $\varepsilon_4$, and the red and green dots are the points on the Weyl surfaces that can be seen in the experiment.

## IV. OBSERVATION OF TOPOLOGICAL PHASE TRANSITION IN FIVE DIMENSIONS

Not only can the circuit designed by us show the above Yang monopole and linked Weyl surfaces, but it also shows the topological phase transition process from the normal insulator to the topological insulator. To observe these phase transitions, we need to adjust the parameters in Eq. (1) as $\eta = 1/2$ [switches turned off in Fig. 1(b)] and $a = 1/2\sqrt{2}$ ($0 \leqslant a < \eta < 1-a$), then change the parameter $m$. When $m < -a$, we find no band intersection in momentum space for such a case, which corresponds to the 5D normal insulator as shown in Fig. 4(a). When $m$ increases to $-a < m < 0$, the Weyl surface between bands $\varepsilon_2$ and $\varepsilon_3$ appears; the loop in Fig. 4(b) represents such a case. Here $m = 1/2 - 1/\sqrt{2}$ is taken. When $m$ increases to $0 < m < a$, another Weyl surface between band $\varepsilon_3$ and $\varepsilon_4$ appears as the green loop in Fig. 4(c), and the previous Weyl surface is stretched longer. The parameter in Fig. 4(c) is $m = 1/2 - 1/2\sqrt{2}$. When $m$ increases to $a < m < 2\eta - a$ (here we choose $m = 1/2$), the Weyl surface between bands $\varepsilon_3$ and $\varepsilon_4$ splits into two smaller Weyl surfaces, and the three Weyl surfaces form a linking structure which can be seen clearly in Fig. 4(d). As $m$ is further increased, the two Weyl surfaces move along $\pm k_z$ and finally merge into a single Weyl surface when $2\eta - a < m < 2\eta + a$ (here we choose $m = 1/2 + 1/2\sqrt{2}$) as shown in Fig. 4(e). The phases in Figs. 4(d) and 4(e) are both linked Weyl surfaces with Weyl arcs on their 4D boundaries just as those we study above. The red Weyl surface then shrinks to zero, and the system becomes a 5D topological insulator for $2\eta + a < m < 2 - a$ (here we choose $m = 1 + 1/2$) as shown in Fig. 4(f). When $m$ keeps increasing until a large positive value, the phase goes from the 5D topological insulator to the normal insulator again. The calculated results based on the circuits are completely consistent with the theoretical results in Ref. [23].

In the circuit experiments, we take different grounding conditions corresponding to variable values of $m$. In Fig. 5 in Appendix A, we show the four grounding conditions for the middle four phases [Figs. 4(b)–4(e)] in this phase transition process. When the first grounding condition is taken, which corresponds to $m = 1/2 - 1/\sqrt{2}$, the agreement between experimental results [red dots in Fig. 4(b)] and theoretical results is very well. When the second grounding condition is taken, experimental results shown by red and green dots in Fig. 4(c) are also identical with the theoretical results. For the third and fourth grounding conditions, the experimental results are described by dots in Figs. 4(d) and 4(e), respectively. Comparing them with the theoretical results, the agreements are observed again. The consistency of experimental and theoretical results shows that a series of phase transitions in 5D Weyl semimetals can be observed in our designed circuit system. Although only some dots from our experiments are provided in Figs. 4(b)–4(e), it is enough to elucidate the evolution of the spectra with the increase of $m$. The change of the intersection between these red and green dots clearly reveals the phase transition in our 5D system. The detailed admittance spectra obtained in our experiments about different phases are shown in Appendix F.

## V. DISCUSSION AND CONCLUSION

We have observed various topological phases and topological phase transitions in 5D Weyl semimetals by constructing 5D electric circuit platforms in fully real space. The 5D Weyl semimetals emerging as intermediate phases during the topological phase transition have also been demonstrated. Because the property of the electric circuit depends only on how sites are connected, regardless of the shape of the circuit lattice, the higher-dimensional electric circuits can always be projected into lower-dimensional spaces with appropriate local/nonlocal site connections. Thus, the circuit networks are regarded to be an ideal platform to study various higher-dimensional physics in real space [40–43]. For example, in elementary particle physics, the spatial dimension is believed to be higher than 3 according to string theory, where these higher-dimensional systems are impossible to approach experimentally in the usual way. Our results will open a route to study these higher-dimensional problems in the laboratory. In addition, our studies also imply unique ways to control electrical signals, and may have potential applications in the field of electronic signal control.

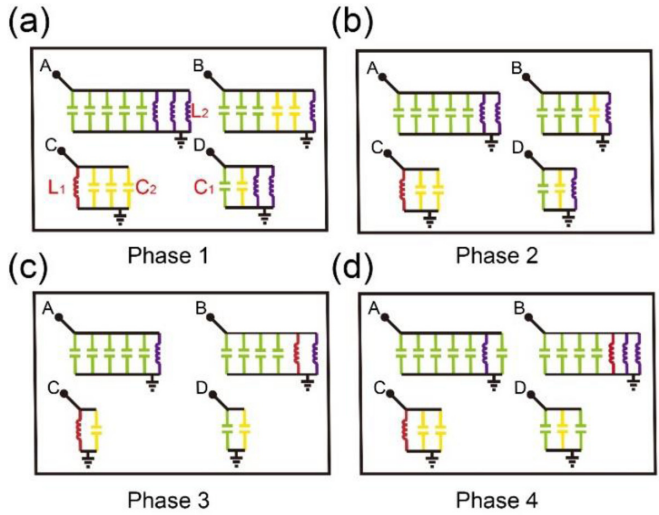

FIG. 5. (a)–(d) show the four grounding conditions in our experiment of phase transition, corresponding to the phases of (b)–(e) of Fig. 4 in the main text.

## ACKNOWLEDGMENTS

We thank Y. Sun for helpful discussions during the research. This work is supported by the National Key R & D Program of China under Grant No. 2017YFA0303800 and the National Natural Science Foundation of China (Grants No. 91850205 and No. 12104041).

## APPENDIX A: DERIVATION OF CIRCUIT LAPLACIAN IN MOMENTUM SPACE FOR 5D WEYL SEMIMETAL CIRCUITS

The designed circuit can be used to exhibit all different phenomena in the 5D Weyl semimetal, which includes the Yang monopole, the linked Weyl surfaces, and 5D topological phase transitions involving the creation of the 5D Weyl semimetal phases. To clearly explain the corresponding relations between the designed circuits and their circuit Laplacians in the experiment, we deduce every circuit Laplacian in the experiment in sequence.

The voltages on sites A–D shown in Figs. 1(b)–1(e) are labeled by $V_A$, $V_B$, $V_C$, and $V_D$, respectively. The corresponding input currents are described by $I_A$, $I_B$, $I_C$, and $I_D$. The phases of the wave vector propagating along the $x$, $y$, $z$, $u$, and $v$ directions are denoted by $k_x$, $k_y$, $k_z$, $k_u$, and $k_v$, respectively. Let us show how to obtain the circuit Laplacian of the Yang monopole first. Now, the switches in Fig. 1(b) are turned off, and the grounding condition is chosen as the Yang monopole [Fig. 1(c)]. After applying Kirchhoff's law to the four nodes, we have

$$
\begin{aligned}
I_A &= \frac{1}{i\omega L_1}(V_A - V_B e^{ik_x}) + \frac{1}{i\omega L_1}(V_A - V_B e^{-ik_x}) + \frac{1}{i\omega L_1}(V_A - V_A e^{ik_y}) + i\omega C_1(V_A - V_B e^{ik_y}) + \frac{1}{i\omega L_1}(V_A - V_A e^{-ik_y}) \\
&\quad + \frac{1}{i\omega L_1}(V_A - V_B e^{-ik_y}) + \frac{1}{i\omega L_1}(V_A - V_A e^{ik_z}) + \frac{1}{i\omega L_1}(V_A - V_A e^{-ik_z}) + \frac{1}{i\omega L_1}(V_A - V_C e^{ik_u}) + \frac{1}{i\omega L_1}(V_A - V_C e^{-ik_u}) \\
&\quad + \frac{1}{i\omega L_1}(V_A - V_A e^{ik_v}) + i\omega C_1(V_A - V_C e^{ik_v}) + \frac{1}{i\omega L_1}(V_A - V_A e^{-ik_v}) + \frac{1}{i\omega L_1}(V_A - V_C e^{-ik_v}) + 6i\omega C_1(V_A - 0),
\end{aligned}
$$

$$
\begin{aligned}
I_B &= \frac{1}{i\omega L_1}(V_B - V_A e^{ik_x}) + \frac{1}{i\omega L_1}(V_B - V_A e^{-ik_x}) + \frac{1}{i\omega L_1}(V_B - V_A e^{ik_y}) + i\omega C_1(V_B - V_B e^{ik_y}) + i\omega C_1(V_B - V_A e^{-ik_y}) \\
&\quad + i\omega C_1(V_B - V_B e^{-ik_y}) + i\omega C_1(V_B - V_B e^{ik_z}) + i\omega C_1(V_B - V_B e^{-ik_z}) + \frac{1}{i\omega L_1}(V_B - V_D e^{ik_u}) + \frac{1}{i\omega L_1}(V_B - V_D e^{ik_u}) \\
&\quad + i\omega C_1(V_B - V_B e^{ik_v}) + i\omega C_1(V_B - V_D e^{ik_v}) + i\omega C_1(V_B - V_B e^{-ik_v}) + \frac{1}{i\omega L_1}(V_B - V_D e^{-ik_v}) + 2i\omega C_1(V_B - 0),
\end{aligned}
$$

$$
\begin{aligned}
I_C &= i\omega C_1(V_C - V_D e^{ik_x}) + i\omega C_1(V_C - V_D e^{-ik_x}) + i\omega C_1(V_C - V_C e^{ik_y}) + \frac{1}{i\omega L_1}(V_C - V_D e^{ik_y}) + i\omega C_1(V_C - V_C e^{-ik_y}) \\
&\quad + i\omega C_1(V_C - V_D e^{-ik_y}) + i\omega C_1(V_C - V_C e^{ik_z}) + i\omega C_1(V_C - V_C e^{-ik_z}) + \frac{1}{i\omega L_1}(V_C - V_A e^{ik_u}) + \frac{1}{i\omega L_1}(V_C - V_A e^{-ik_u}) \\
&\quad + \frac{1}{i\omega L_1}(V_C - V_A e^{ik_v}) + i\omega C_1(V_C - V_C e^{ik_v}) + i\omega C_1(V_C - V_A e^{-ik_v}) + i\omega C_1(V_C - V_C e^{-ik_v}) + \frac{2}{i\omega L_1}(V_C - 0),
\end{aligned}
$$

$$
\begin{aligned}
I_D &= i\omega C_1(V_D - V_C e^{ik_x}) + i\omega C_1(V_D - V_C e^{-ik_x}) + i\omega C_1(V_D - V_C e^{ik_y}) + \frac{1}{i\omega L_1}(V_D - V_D e^{ik_y}) + \frac{1}{i\omega L_1}(V_D - V_C e^{ik_y}) \\
&\quad + \frac{1}{i\omega L_1}(V_D - V_D e^{-ik_y}) + \frac{1}{i\omega L_1}(V_D - V_D e^{ik_z}) + \frac{1}{i\omega L_1}(V_D - V_D e^{-ik_z}) + \frac{1}{i\omega L_1}(V_D - V_C e^{ik_u}) + \frac{1}{i\omega L_1}(V_D - V_C e^{-ik_u}) \\
&\quad + \frac{1}{i\omega L_1}(V_D - V_B e^{ik_v}) + \frac{1}{i\omega L_1}(V_D - V_D e^{ik_v}) + i\omega C_1(V_D - V_B e^{-ik_v}) + \frac{1}{i\omega L_1}(V_D - V_D e^{-ik_v}) + 2i\omega C_1(V_D - 0). \quad \text{(A1)}
\end{aligned}
$$

Equation (A1) can be written in the form of $\mathbf{I} = J\mathbf{V}$. When the driving frequency satisfies $\omega = \omega_0 = 1/\sqrt{L_1C_1}$, we have $-1/(i\omega L_1) = i\omega C_1 = i\sqrt{C_1/L_1}$. The circuit Laplacian in momentum space can be expressed as

$$
J = i\sqrt{\frac{C_1}{L_1}}\begin{pmatrix}
e^{ik_x}+e^{-ik_x}+e^{ik_y}+e^{-ik_y}+e^{ik_z}+e^{-ik_z}-4 & e^{ik_x}+e^{-ik_x}-e^{ik_y}+e^{-ik_y} & e^{ik_x}+e^{-ik_x}-e^{ik_y}+e^{-ik_z} & 0 \\
e^{ik_x}+e^{-ik_x}+e^{ik_y}-e^{-ik_z} & -e^{ik_x}-e^{-ik_x}-e^{ik_y}-e^{-ik_z}-e^{ik_x}-e^{-ik_z}+4 & 0 & e^{ik_x}+e^{-ik_x}-e^{ik_y}+e^{-ik_z} \\
e^{ik_x}+e^{-ik_x}+e^{ik_y}-e^{-ik_z} & 0 & -e^{ik_x}-e^{-ik_x}-e^{ik_y}-e^{-ik_z}-e^{ik_x}-e^{-ik_z}+4 & -e^{ik_x}-e^{-ik_x}+e^{ik_y}-e^{-ik_z} \\
0 & e^{ik_x}+e^{-ik_x}+e^{ik_y}-e^{-ik_z} & -e^{ik_x}-e^{-ik_x}-e^{ik_y}+e^{-ik_z} & e^{ik_x}+e^{-ik_x}+e^{ik_y}+e^{-ik_y}+e^{ik_z}+e^{-ik_z}-4
\end{pmatrix}.
$$
(A2)

We set $\Gamma^{1,2,3,4,5} = \{\sigma_3\tau_1, \sigma_3\tau_2, \sigma_3\tau_3, \sigma_1, \sigma_2\}$, where $\sigma$ and $\tau$ are both Pauli matrices, and define the functions $\xi_i(\mathbf{k})$ as $\zeta_1(\mathbf{k}) = \cos k_x$, $\zeta_2(\mathbf{k}) = \sin k_y$, $\zeta_3(\mathbf{k}) = \eta(\cos k_z - 1) + (\cos k_y + \cos k_v - 2) + m$, $\zeta_4(\mathbf{k}) = \cos k_u$, and $\zeta_5(\mathbf{k}) = \sin k_v$. In this case, Eq. (A2) can be transformed to Eq. (1) in the main text with $a = 0$, $\eta = 1$, and $m = 1$. These are the parameters of the Yang monopole system.

For the linked Weyl surface system, the switches in Fig. 1(b) are turned off, and the grounding condition is chosen as the linked Weyl surfaces shown in Fig. 1(d). After applying Kirchhoff's law to the four nodes, we have

$$
\begin{aligned}
I_A = &\frac{1}{i\omega L_1}(V_A - V_B e^{ik_x}) + \frac{1}{i\omega L_1}(V_A - V_B e^{-ik_x}) + \frac{1}{i\omega L_1}(V_A - V_A e^{ik_y}) + i\omega C_1(V_A - V_B e^{ik_y}) + \frac{1}{i\omega L_1}(V_A - V_A e^{-ik_y}) \\
&+ \frac{1}{i\omega L_1}(V_A - V_B e^{-ik_y}) + \frac{1}{i\omega L_1}(V_A - V_A e^{ik_z}) + \frac{1}{i\omega L_1}(V_A - V_A e^{-ik_z}) + \frac{1}{i\omega L_1}(V_A - V_C e^{ik_u}) + \frac{1}{i\omega L_1}(V_A - V_C e^{-ik_u}) \\
&+ \frac{1}{i\omega L_1}(V_A - V_A e^{ik_v}) + i\omega C_1(V_A - V_C e^{ik_v}) + \frac{1}{i\omega L_1}(V_A - V_A e^{-ik_v}) + \frac{1}{i\omega L_1}(V_A - V_C e^{-ik_v}) + 4i\omega C_1(V_A - 0),
\end{aligned}
$$

$$
\begin{aligned}
I_B = &\frac{1}{i\omega L_1}(V_B - V_A e^{ik_x}) + \frac{1}{i\omega L_1}(V_B - V_A e^{-ik_x}) + \frac{1}{i\omega L_1}(V_B - V_A e^{ik_y}) + i\omega C_1(V_B - V_B e^{ik_y}) + i\omega C_1(V_B - V_A e^{-ik_y}) \\
&+ i\omega C_1(V_B - V_B e^{-ik_y}) + i\omega C_1(V_B - V_B e^{ik_z}) + i\omega C_1(V_B - V_B e^{-ik_z}) + \frac{1}{i\omega L_1}(V_B - V_D e^{ik_u}) + \frac{1}{i\omega L_1}(V_B - V_D e^{-ik_u}) \\
&+ i\omega C_1(V_B - V_B e^{ik_v}) + i\omega C_1(V_B - V_D e^{ik_v}) + i\omega C_1(V_B - V_B e^{-ik_v}) + \frac{1}{i\omega L_1}(V_B - V_D e^{-ik_v}) + i\omega C_1(V_B - 0) + \frac{1}{i\omega L_1}(V_B - 0),
\end{aligned}
$$

$$
\begin{aligned}
I_C = &i\omega C_1(V_C - V_D e^{ik_x}) + i\omega C_1(V_C - V_D e^{-ik_x}) + i\omega C_1(V_C - V_C e^{ik_y}) + i\omega C_1(V_C - V_D e^{ik_y}) + i\omega C_1(V_C - V_C e^{-ik_y}) \\
&+ i\omega C_1(V_C - V_D e^{-ik_y}) + i\omega C_1(V_C - V_C e^{ik_z}) + i\omega C_1(V_C - V_C e^{-ik_z}) + \frac{1}{i\omega L_1}(V_C - V_A e^{ik_u}) + \frac{1}{i\omega L_1}(V_C - V_A e^{-ik_u}) \\
&+ \frac{1}{i\omega L_1}(V_C - V_A e^{ik_v}) + i\omega C_1(V_C - V_C e^{ik_v}) + i\omega C_1(V_C - V_A e^{-ik_v}) + i\omega C_1(V_C - V_C e^{-ik_v}) + \frac{2}{i\omega L_1}(V_C - 0) + 2i\omega C_2(V_C - 0),
\end{aligned}
$$

$$
\begin{aligned}
I_D = &i\omega C_1(V_D - V_C e^{ik_x}) + i\omega C_1(V_D - V_C e^{-ik_x}) + i\omega C_1(V_D - V_C e^{ik_y}) + \frac{1}{i\omega L_1}(V_D - V_D e^{ik_y}) + \frac{1}{i\omega L_1}(V_D - V_C e^{-ik_y}) \\
&+ \frac{1}{i\omega L_1}(V_D - V_D e^{-ik_y}) + \frac{1}{i\omega L_1}(V_D - V_D e^{ik_z}) + \frac{1}{i\omega L_1}(V_D - V_D e^{-ik_z}) + \frac{1}{i\omega L_1}(V_D - V_C e^{ik_u}) + \frac{1}{i\omega L_1}(V_D - V_C e^{-ik_u}) \\
&+ \frac{1}{i\omega L_1}(V_D - V_B e^{ik_v}) + \frac{1}{i\omega L_1}(V_D - V_D e^{ik_v}) + i\omega C_1(V_D - V_B e^{-ik_v}) + \frac{1}{i\omega L_1}(V_D - V_D e^{-ik_v}) \\
&+ 2i\omega C_1(V_D - 0) + 2i\omega C_2(V_D - 0).
\end{aligned}
$$
(A3)

Equation (A3) can be written in the form of $\mathbf{I} = J\mathbf{V}$. When the driving frequency is set as $\omega = \omega_0 = 1/\sqrt{L_1C_1} = 1/\sqrt{L_2C_2}$, the circuit Laplacian in momentum space can be written as

$$
J = i\sqrt{\frac{C_1}{L_1}}\begin{pmatrix}
e^{ik_x}+e^{-ik_x}+e^{ik_y}+e^{-ik_y}+e^{ik_z}+e^{-ik_z}-4-\sqrt{2} & e^{ik_x}+e^{-ik_x}-e^{ik_y}+e^{-ik_y} & e^{ik_x}+e^{-ik_x}-e^{ik_y}+e^{-ik_z} & 0 \\
e^{ik_x}+e^{-ik_x}+e^{ik_y}-e^{-ik_z} & -e^{ik_x}-e^{-ik_x}-e^{ik_y}-e^{-ik_z}-e^{ik_x}-e^{-ik_z}+4-\sqrt{2} & 0 & e^{ik_x}+e^{-ik_x}-e^{ik_y}+e^{-ik_z} \\
e^{ik_x}+e^{-ik_x}+e^{ik_y}-e^{-ik_z} & 0 & -e^{ik_x}-e^{-ik_x}-e^{ik_y}-e^{-ik_z}-e^{ik_x}-e^{-ik_z}+4+\sqrt{2} & -e^{ik_x}-e^{-ik_x}+e^{ik_y}-e^{-ik_z} \\
0 & e^{ik_x}+e^{-ik_x}+e^{ik_y}-e^{-ik_z} & -e^{ik_x}-e^{-ik_x}-e^{ik_y}+e^{-ik_z} & e^{ik_x}+e^{-ik_x}+e^{ik_y}+e^{-ik_y}+e^{ik_z}+e^{-ik_z}-4+\sqrt{2}
\end{pmatrix}.
$$
(A4)

which is exactly the same as Eq. (1) in the main text when $a = 1/\sqrt{2}$, $\eta = 1$, and $m = 1$. They are the parameters of the system with the linked Weyl surfaces.

In the discussion of the topological phase transition in 5D of our work, we experimentally realize four different topological phases, which have been shown as Figs. 4(b) and 4(c) in the main text. Here the four phases are marked as phases 1–4. They differ only in grounding conditions which are shown in Fig. 5. For the phase 1, the switches in Fig. 1(b) are turned on, and the grounding condition is chosen as Fig. 5(a). After applying Kirchhoff's law to the four nodes, we have

$$
\begin{aligned}
I_A &= \frac{1}{i\omega L_1}(V_A - V_B e^{ik_x}) + \frac{1}{i\omega L_1}(V_A - V_B e^{-ik_x}) + \frac{1}{i\omega L_1}(V_A - V_A e^{ik_y}) + i\omega C_1(V_A - V_B e^{ik_y}) + \frac{1}{i\omega L_1}(V_A - V_A e^{-ik_y}) \\
&\quad + \frac{1}{i\omega L_1}(V_A - V_B e^{-ik_y}) + \frac{1}{2i\omega L_1}(V_A - V_A e^{ik_z}) + \frac{1}{2i\omega L_1}(V_A - V_A e^{-ik_z}) + \frac{1}{i\omega L_1}(V_A - V_C e^{ik_u}) + \frac{1}{i\omega L_1}(V_A - V_C e^{-ik_u}) \\
&\quad + \frac{1}{i\omega L_1}(V_A - V_A e^{ik_v}) + i\omega C_1(V_A - V_C e^{ik_v}) + \frac{1}{i\omega L_1}(V_A - V_A e^{-ik_v}) + \frac{1}{i\omega L_1}(V_A - V_C e^{-ik_v}) + 5i\omega C_1(V_A - 0) + \frac{3}{i\omega L_2}(V_A - 0),
\end{aligned}
$$

$$
\begin{aligned}
I_B &= \frac{1}{i\omega L_1}(V_B - V_A e^{ik_x}) + \frac{1}{i\omega L_1}(V_B - V_A e^{-ik_x}) + \frac{1}{i\omega L_1}(V_B - V_A e^{ik_y}) + i\omega C_1(V_B - V_B e^{ik_y}) + i\omega C_1(V_B - V_A e^{-ik_y}) \\
&\quad + i\omega C_1(V_B - V_B e^{-ik_y}) + \frac{i\omega C_1}{2}(V_B - V_B e^{ik_z}) + \frac{i\omega C_1}{2}(V_B - V_B e^{-ik_z}) + \frac{1}{i\omega L_1}(V_B - V_D e^{ik_u}) + \frac{1}{i\omega L_1}(V_B - V_D e^{-ik_u}) \\
&\quad + i\omega C_1(V_B - V_B e^{ik_v}) + i\omega C_1(V_B - V_D e^{ik_v}) + i\omega C_1(V_B - V_B e^{-ik_v}) + \frac{1}{i\omega L_1}(V_B - V_D e^{-ik_v}) \\
&\quad + 3i\omega C_1(V_B - 0) + 2i\omega C_2(V_B - 0) + \frac{1}{i\omega L_2}(V_B - 0),
\end{aligned}
$$

$$
\begin{aligned}
I_C &= i\omega C_1(V_C - V_D e^{ik_x}) + i\omega C_1(V_C - V_D e^{-ik_x}) + i\omega C_1(V_C - V_C e^{ik_y}) + \frac{1}{i\omega L_1}(V_C - V_D e^{ik_y}) + i\omega C_1(V_C - V_C e^{-ik_y}) \\
&\quad + i\omega C_1(V_C - V_D e^{-ik_y}) + \frac{i\omega C_1}{2}(V_C - V_C e^{ik_z}) + \frac{i\omega C_1}{2}(V_C - V_C e^{-ik_z}) + \frac{1}{i\omega L_1}(V_C - V_A e^{ik_u}) + \frac{1}{i\omega L_1}(V_C - V_A e^{-ik_u}) \\
&\quad + \frac{1}{i\omega L_1}(V_C - V_A e^{ik_v}) + i\omega C_1(V_C - V_C e^{ik_v}) + i\omega C_1(V_C - V_A e^{-ik_v}) + i\omega C_1(V_C - V_C e^{-ik_v}) + \frac{1}{i\omega L_1}(V_C - 0) + 3i\omega C_2(V_C - 0),
\end{aligned}
$$

$$
\begin{aligned}
I_D &= i\omega C_1(V_D - V_C e^{ik_x}) + i\omega C_1(V_D - V_C e^{-ik_x}) + i\omega C_1(V_D - V_C e^{ik_y}) + \frac{1}{i\omega L_1}(V_D - V_D e^{ik_y}) + \frac{1}{i\omega L_1}(V_D - V_C e^{-ik_y}) \\
&\quad + \frac{1}{i\omega L_1}(V_D - V_D e^{-ik_y}) + \frac{1}{2i\omega L_1}(V_D - V_D e^{ik_z}) + \frac{1}{2i\omega L_1}(V_D - V_D e^{-ik_z}) \\
&\quad + \frac{1}{i\omega L_1}(V_D - V_C e^{ik_u}) + \frac{1}{i\omega L_1}(V_D - V_C e^{-ik_u}) + \frac{1}{i\omega L_1}(V_D - V_B e^{ik_v}) + \frac{1}{i\omega L_1}(V_D - V_D e^{ik_v}) + i\omega C_1(V_D - V_B e^{-ik_v}) \\
&\quad + \frac{1}{i\omega L_1}(V_D - V_D e^{-ik_v}) + i\omega C_1(V_D - 0) + i\omega C_2(V_D - 0) + \frac{2}{i\omega L_2}(V_D - 0).
\end{aligned}
\tag{A5}
$$

Equation (A5) can be written in the form of $\mathbf{I} = J\mathbf{V}$. When the driving frequency is set as $\omega = \omega_0$, the circuit Laplacian in momentum space can be written as

$$
J = i\sqrt{\frac{C_1}{L_1}}
\begin{pmatrix}
e^{ik_x}+e^{-ik_x}+e^{ik_y}+e^{-ik_y}+e^{ik_v}+e^{-ik_v}-4-\frac{3}{\sqrt{2}} & e^{ik_y}+e^{-ik_y}-e^{ik_x}+e^{-ik_v} & e^{ik_u}+e^{-ik_u}-e^{ik_v}+e^{-ik_v} & 0 \\
e^{ik_x}+e^{-ik_x}+e^{ik_y}-e^{-ik_v} & -e^{ik_y}-e^{-ik_y}-e^{ik_z}-e^{-ik_z}-e^{ik_v}+4+\frac{1}{\sqrt{2}} & 0 & e^{ik_u}+e^{-ik_u}-e^{ik_v}+e^{-ik_v} \\
e^{ik_v}+e^{-ik_v}+e^{ik_u}-e^{-ik_u} & 0 & -e^{ik_y}-e^{-ik_y}-e^{ik_z}-e^{-ik_z}-e^{ik_v}+4+\frac{1}{\sqrt{2}} & -e^{ik_y}-e^{-ik_y}+e^{ik_v}-e^{-ik_v} \\
0 & e^{ik_v}+e^{-ik_v}+e^{ik_u}-e^{-ik_u} & -e^{ik_x}-e^{-ik_x}-e^{ik_v}+e^{-ik_v} & e^{ik_v}+e^{-ik_v}+e^{ik_x}+e^{-ik_x}+e^{ik_u}+e^{-ik_u}-4-\frac{3}{\sqrt{2}}
\end{pmatrix}.
\tag{A6}
$$

Here, it is exactly the same as Eq. (1) in the main text when $a = 1/\sqrt{8}$, $\eta = 1/2$, and $m = 1/2 - 1/\sqrt{2}$.

For phase 2 in the 5D system, the switches in Fig. 1(b) are turned on and the grounding condition is chosen as Fig. 5(b). After applying Kirchhoff's law to the four nodes, we have

$$
\begin{aligned}
I_A &= \frac{1}{i\omega L_1}(V_A - V_B e^{ik_x}) + \frac{1}{i\omega L_1}(V_A - V_B e^{-ik_x}) + \frac{1}{i\omega L_1}(V_A - V_A e^{ik_y}) + i\omega C_1(V_A - V_B e^{ik_y}) + \frac{1}{i\omega L_1}(V_A - V_A e^{-ik_y}) \\
&\quad + \frac{1}{i\omega L_1}(V_A - V_B e^{-ik_y}) + \frac{1}{2i\omega L_1}(V_A - V_A e^{ik_z}) + \frac{1}{2i\omega L_1}(V_A - V_A e^{-ik_z}) + \frac{1}{i\omega L_1}(V_A - V_C e^{ik_u}) + \frac{1}{i\omega L_1}(V_A - V_C e^{-ik_u}) \\
&\quad + \frac{1}{i\omega L_1}(V_A - V_A e^{ik_v}) + i\omega C_1(V_A - V_C e^{ik_v}) + \frac{1}{i\omega L_1}(V_A - V_A e^{-ik_v}) + \frac{1}{i\omega L_1}(V_A - V_C e^{-ik_v}) + 5i\omega C_1(V_A - 0) + \frac{2}{i\omega L_2}(V_A - 0),
\end{aligned}
$$

$$I_B = \frac{1}{i\omega L_1}(V_B - V_A e^{ik_x}) + \frac{1}{i\omega L_1}(V_B - V_A e^{-ik_x}) + \frac{1}{i\omega L_1}(V_B - V_A e^{ik_y}) + i\omega C_1(V_B - V_B e^{ik_y}) + i\omega C_1(V_B - V_A e^{-ik_y})$$

$$+ i\omega C_1(V_B - V_B e^{-ik_y}) + \frac{i\omega C_1}{2}(V_B - V_B e^{ik_z}) + \frac{i\omega C_1}{2}(V_B - V_B e^{-ik_z}) + \frac{1}{i\omega L_1}(V_B - V_D e^{ik_u}) + \frac{1}{i\omega L_1}(V_B - V_D e^{-ik_u})$$

$$+ i\omega C_1(V_B - V_B e^{ik_v}) + i\omega C_1(V_B - V_D e^{-ik_v}) + i\omega C_1(V_B - V_B e^{-ik_v}) + \frac{1}{i\omega L_1}(V_B - V_D e^{-ik_v}) + 3i\omega C_1(V_B - 0)$$

$$+ i\omega C_2(V_B - 0) + \frac{1}{i\omega L_2}(V_B - 0),$$

$$I_C = i\omega C_1(V_C - V_D e^{ik_x}) + i\omega C_1(V_C - V_D e^{-ik_x}) + i\omega C_1(V_C - V_C e^{ik_y}) + \frac{1}{i\omega L_1}(V_C - V_D e^{ik_y}) + i\omega C_1(V_C - V_C e^{-ik_y})$$

$$+ i\omega C_1(V_C - V_D e^{-ik_y}) + \frac{i\omega C_1}{2}(V_C - V_C e^{ik_z}) + \frac{i\omega C_1}{2}(V_C - V_C e^{-ik_z}) + \frac{1}{i\omega L_1}(V_C - V_A e^{ik_u}) + \frac{1}{i\omega L_1}(V_C - V_A e^{-ik_u})$$

$$+ \frac{1}{i\omega L_1}(V_C - V_A e^{ik_v}) + i\omega C_1(V_C - V_C e^{ik_v}) + i\omega C_1(V_C - V_A e^{-ik_v}) + i\omega C_1(V_C - V_C e^{-ik_v}) + \frac{1}{i\omega L_1}(V_C - 0) + 2i\omega C_2(V_C - 0),$$

$$I_D = i\omega C_1(V_D - V_C e^{ik_x}) + i\omega C_1(V_D - V_C e^{-ik_x}) + i\omega C_1(V_D - V_C e^{ik_y}) + \frac{1}{i\omega L_1}(V_D - V_D e^{ik_y}) + \frac{1}{i\omega L_1}(V_D - V_C e^{-ik_y})$$

$$+ \frac{1}{i\omega L_1}(V_D - V_D e^{-iky}) + \frac{1}{2i\omega L_1}(V_D - V_D e^{ik_z}) + \frac{1}{2i\omega L_1}(V_D - V_D e^{-ik_z}) + \frac{1}{i\omega L_1}(V_D - V_C e^{ik_u}) + \frac{1}{i\omega L_1}(V_D - V_C e^{-ik_u})$$

$$+ \frac{1}{i\omega L_1}(V_D - V_B e^{ik_v}) + \frac{1}{i\omega L_1}(V_D - V_D e^{ik_v}) + i\omega C_1(V_D - V_B e^{-ik_v}) + \frac{1}{i\omega L_1}(V_D - V_D e^{-ik_v})$$

$$+ i\omega C_1(V_D - 0) + i\omega C_2(V_D - 0) + \frac{1}{i\omega L_2}(V_D - 0). \tag{A7}$$

Equation (A7) can be written in the form of $\mathbf{I} = J\mathbf{V}$. When the driving frequency is set as $\omega = \omega_0$, the circuit Laplacian in momentum space can be written as

$$J = i\sqrt{\frac{C_1}{L_1}}\begin{pmatrix} e^{ik_y} + e^{-ik_y} + e^{ik_z} + e^{-ik_z} + e^{ik_v} + e^{-ik_v} - 4 - \sqrt{2} & e^{ik_y} + e^{-ik_y} - e^{ik_v} + e^{-ik_v} & e^{ik_y} + e^{-ik_y} - e^{ik_v} + e^{-ik_v} & 0 \\ e^{ik_y} + e^{-ik_y} + e^{ik_v} - e^{-ik_v} & -e^{ik_y} - e^{-ik_y} - e^{ik_v} - e^{-ik_v} - e^{ik_z} - e^{-ik_z} + 4 & 0 & e^{ik_y} + e^{-ik_y} - e^{ik_v} + e^{-ik_v} \\ e^{ik_x} + e^{-ik_x} + e^{ik_v} - e^{-ik_v} & 0 & -e^{ik_y} - e^{-ik_y} - e^{ik_v} - e^{-ik_v} - e^{ik_z} - e^{-ik_z} + 4 + \sqrt{2} & -e^{ik_x} - e^{-ik_x} - e^{ik_v} + e^{-ik_v} \\ 0 & e^{ik_x} + e^{-ik_x} + e^{ik_v} - e^{-ik_v} & -e^{ik_x} - e^{-ik_x} - e^{ik_v} + e^{-ik_v} & e^{ik_y} + e^{-ik_y} + e^{ik_z} + e^{-ik_z} + e^{ik_v} + e^{-ik_v} - 4 \end{pmatrix} \tag{A8}$$

It is exactly the same as Eq. (1) in the main text when $a = 1/\sqrt{8}$, $\eta = 1/2$, and $m = 1/2 - 1/(2\sqrt{2})$.

For phase 3 in the 5D system, the switches in Fig. 1(b) are turned on, and the grounding condition is chosen as Fig. 5(c). After applying Kirchhoff's law to the four nodes, we have

$$I_A = \frac{1}{i\omega L_1}(V_A - V_B e^{ik_x}) + \frac{1}{i\omega L_1}(V_A - V_B e^{-ik_x}) + \frac{1}{i\omega L_1}(V_A - V_A e^{ik_y}) + i\omega C_1(V_A - V_B e^{ik_y}) + \frac{1}{i\omega L_1}(V_A - V_A e^{-ik_y})$$

$$+ \frac{1}{i\omega L_1}(V_A - V_B e^{-ik_y}) + \frac{1}{2i\omega L_1}(V_A - V_A e^{ik_z}) + \frac{1}{2i\omega L_1}(V_A - V_A e^{-ik_z}) + \frac{1}{i\omega L_1}(V_A - V_C e^{ik_u}) + \frac{1}{i\omega L_1}(V_A - V_C e^{-ik_u})$$

$$+ \frac{1}{i\omega L_1}(V_A - V_A e^{ik_v}) + i\omega C_1(V_A - V_C e^{ik_v}) + \frac{1}{i\omega L_1}(V_A - V_A e^{-ik_v}) + \frac{1}{i\omega L_1}(V_A - V_C e^{-ik_v}) + 5i\omega C_1(V_A - 0) + \frac{1}{i\omega L_2}(V_A - 0),$$

$$I_B = \frac{1}{i\omega L_1}(V_B - V_A e^{ik_x}) + \frac{1}{i\omega L_1}(V_B - V_A e^{-ik_x}) + \frac{1}{i\omega L_1}(V_B - V_A e^{ik_y}) + i\omega C_1(V_B - V_B e^{ik_y}) + i\omega C_1(V_B - V_A e^{-ik_y})$$

$$+ i\omega C_1(V_B - V_B e^{-ik_y}) + \frac{i\omega C_1}{2}(V_B - V_B e^{ik_z}) + \frac{i\omega C_1}{2}(V_B - V_B e^{-ik_z}) + \frac{1}{i\omega L_1}(V_B - V_D e^{ik_u}) + \frac{1}{i\omega L_1}(V_B - V_D e^{-ik_u})$$

$$+ i\omega C_1(V_B - V_B e^{ik_v}) + i\omega C_1(V_B - V_D e^{-ik_v}) + i\omega C_1(V_B - V_B e^{-ik_v}) + \frac{1}{i\omega L_1}(V_B - V_D e^{-ik_v})$$

$$+ 4i\omega C_1(V_B - 0) + \frac{1}{i\omega L_1}(V_B - 0) + \frac{1}{i\omega L_2}(V_B - 0),$$

$$I_C = i\omega C_1(V_C - V_D e^{ik_x}) + i\omega C_1(V_C - V_D e^{-ik_x}) + i\omega C_1(V_C - V_C e^{ik_y}) + \frac{1}{i\omega L_1}(V_C - V_D e^{ik_y}) + i\omega C_1(V_C - V_C e^{-ik_y})$$

$$+ i\omega C_1(V_C - V_D e^{-ik_y}) + \frac{i\omega C_1}{2}(V_C - V_C e^{ik_z}) + \frac{i\omega C_1}{2}(V_C - V_C e^{-ik_z}) + \frac{1}{i\omega L_1}(V_C - V_A e^{ik_u}) + \frac{1}{i\omega L_1}(V_C - V_A e^{-ik_u})$$

$$+ \frac{1}{i\omega L_1}(V_C - V_A e^{ik_v}) + i\omega C_1(V_C - V_C e^{ik_v}) + i\omega C_1(V_C - V_A e^{-ik_v}) + i\omega C_1(V_C - V_C e^{-ik_v}) + \frac{1}{i\omega L_1}(V_C - 0) + i\omega C_2(V_C - 0),$$

$$
\begin{aligned}
I_D &= i\omega C_1(V_D - V_C e^{ik_x}) + i\omega C_1(V_D - V_C e^{-ik_x}) + i\omega C_1(V_D - V_C e^{ik_y}) + \frac{1}{i\omega L_1}(V_D - V_D e^{ik_y}) + \frac{1}{i\omega L_1}(V_D - V_C e^{-ik_y}) \\
&\quad + \frac{1}{i\omega L_1}(V_D - V_D e^{-ik_y}) + \frac{1}{2i\omega L_1}(V_D - V_D e^{ik_z}) + \frac{1}{2i\omega L_1}(V_D - V_D e^{-ik_z}) \\
&\quad + \frac{1}{i\omega L_1}(V_D - V_C e^{ik_u}) + \frac{1}{i\omega L_1}(V_D - V_C e^{-ik_u}) + \frac{1}{i\omega L_1}(V_D - V_B e^{ik_v}) + \frac{1}{i\omega L_1}(V_D - V_D e^{ik_v}) + i\omega C_1(V_D - V_B e^{-ik_v}) \\
&\quad + \frac{1}{i\omega L_1}(V_D - V_D e^{-ik_v}) + i\omega C_1(V_D - 0) + i\omega C_2(V_D - 0).
\end{aligned} \tag{A9}
$$

Equation (A9) can be written in the form of $\mathbf{I} = J\mathbf{V}$. When the driving frequency is set as $\omega = \omega_0$, the circuit Laplacian in momentum space can be written as

$$
J = i\sqrt{\frac{C_1}{L_1}} \begin{pmatrix}
e^{ik_x} + e^{-ik_x} + e^{ik_y} + e^{-ik_y} + e^{ik_v} + e^{-ik_v} - 4 - \frac{1}{\sqrt{2}} & e^{ik_v} + e^{-ik_v} - e^{ik_x} + e^{-ik_x} & e^{ik_y} + e^{-ik_y} - e^{ik_u} + e^{-ik_u} & 0 \\
e^{ik_v} + e^{-ik_v} + e^{ik_x} - e^{-ik_x} & -e^{ik_v} - e^{-ik_v} - e^{ik_x} - e^{-ik_x} - e^{ik_u} - e^{-ik_u} + 4 - \frac{1}{\sqrt{2}} & 0 & e^{ik_u} + e^{-ik_u} - e^{ik_x} + e^{-ik_x} \\
e^{ik_y} + e^{-ik_y} + e^{ik_u} - e^{-ik_u} & 0 & -e^{ik_y} - e^{-ik_y} - e^{ik_u} - e^{-ik_u} - e^{ik_x} - e^{-ik_x} + 4 + \frac{1}{\sqrt{2}} & -e^{ik_x} - e^{-ik_x} + e^{ik_v} - e^{-ik_v} \\
0 & e^{ik_u} + e^{-ik_u} + e^{ik_x} - e^{-ik_x} & e^{ik_v} + e^{-ik_v} + e^{ik_x} - e^{-ik_x} & e^{ik_x} + e^{-ik_x} + e^{ik_u} + e^{-ik_u} + e^{-ik_v} + e^{ik_v} - 4 + \frac{1}{\sqrt{2}}
\end{pmatrix}. \tag{A10}
$$

In this case, it is exactly the same as Eq. (1) in the main text when $a = 1/\sqrt{8}$, $\eta = 1/2$, and $m = 1/2$.

For phase 4 in the 5D system, the switches in Fig. 1(b) are turned on, and the grounding condition is chosen as Fig. 5(d). After applying Kirchhoff's law to the four nodes, we have

$$
\begin{aligned}
I_A &= \frac{1}{i\omega L_1}(V_A - V_B e^{ik_x}) + \frac{1}{i\omega L_1}(V_A - V_B e^{-ik_x}) + \frac{1}{i\omega L_1}(V_A - V_A e^{ik_y}) + i\omega C_1(V_A - V_B e^{ik_y}) + \frac{1}{i\omega L_1}(V_A - V_A e^{-ik_y}) \\
&\quad + \frac{1}{i\omega L_1}(V_A - V_B e^{-ik_y}) + \frac{1}{2i\omega L_1}(V_A - V_A e^{ik_z}) + \frac{1}{2i\omega L_1}(V_A - V_A e^{-ik_z}) + \frac{1}{i\omega L_1}(V_A - V_C e^{ik_u}) + \frac{1}{i\omega L_1}(V_A - V_C e^{-ik_u}) \\
&\quad + \frac{1}{i\omega L_1}(V_A - V_A e^{ik_v}) + i\omega C_1(V_A - V_C e^{ik_v}) + \frac{1}{i\omega L_1}(V_A - V_A e^{-ik_v}) + \frac{1}{i\omega L_1}(V_A - V_C e^{-ik_v}) + 6i\omega C_1(V_A - 0) + \frac{1}{i\omega L_2}(V_A - 0),
\end{aligned}
$$

$$
\begin{aligned}
I_B &= \frac{1}{i\omega L_1}(V_B - V_A e^{ik_x}) + \frac{1}{i\omega L_1}(V_B - V_A e^{-ik_x}) + \frac{1}{i\omega L_1}(V_B - V_A e^{ik_y}) + i\omega C_1(V_B - V_B e^{ik_y}) + i\omega C_1(V_B - V_A e^{-ik_y}) \\
&\quad + i\omega C_1(V_B - V_B e^{-ik_y}) + \frac{i\omega C_1}{2}(V_B - V_B e^{ik_z}) + \frac{i\omega C_1}{2}(V_B - V_B e^{-ik_z}) + \frac{1}{i\omega L_1}(V_B - V_D e^{ik_u}) + \frac{1}{i\omega L_1}(V_B - V_D e^{-ik_u}) \\
&\quad + i\omega C_1(V_B - V_B e^{ik_v}) + i\omega C_1(V_B - V_D e^{ik_v}) + i\omega C_1(V_B - V_B e^{-ik_v}) + \frac{1}{i\omega L_1}(V_B - V_D e^{-ik_v}) \\
&\quad + 4i\omega C_1(V_B - 0) + \frac{1}{i\omega L_1}(V_B - 0) + \frac{2}{i\omega L_2}(V_B - 0),
\end{aligned}
$$

$$
\begin{aligned}
I_C &= i\omega C_1(V_C - V_D e^{ik_x}) + i\omega C_1(V_C - V_D e^{-ik_x}) + i\omega C_1(V_C - V_C e^{ik_y}) + \frac{1}{i\omega L_1}(V_C - V_D e^{ik_y}) + i\omega C_1(V_C - V_C e^{-ik_y}) \\
&\quad + i\omega C_1(V_C - V_D e^{-ik_y}) + \frac{i\omega C_1}{2}(V_C - V_C e^{ik_z}) + \frac{i\omega C_1}{2}(V_C - V_C e^{-ik_z}) + \frac{1}{i\omega L_1}(V_C - V_A e^{ik_u}) + \frac{1}{i\omega L_1}(V_C - V_A e^{-ik_u}) \\
&\quad + \frac{1}{i\omega L_1}(V_C - V_A e^{ik_v}) + i\omega C_1(V_C - V_C e^{ik_v}) + i\omega C_1(V_C - V_A e^{-ik_v}) + i\omega C_1(V_C - V_C e^{-ik_v}) + \frac{1}{i\omega L_1}(V_C - 0) + 2i\omega C_2(V_C - 0),
\end{aligned}
$$

$$
\begin{aligned}
I_D &= i\omega C_1(V_D - V_C e^{ik_x}) + i\omega C_1(V_D - V_C e^{-ik_x}) + i\omega C_1(V_D - V_C e^{ik_y}) + \frac{1}{i\omega L_1}(V_D - V_D e^{ik_y}) + \frac{1}{i\omega L_1}(V_D - V_C e^{-ik_y}) \\
&\quad + \frac{1}{i\omega L_1}(V_D - V_D e^{-ik_y}) + \frac{1}{2i\omega L_1}(V_D - V_D e^{ik_z}) + \frac{1}{2i\omega L_1}(V_D - V_D e^{-ik_z}) + \frac{1}{i\omega L_1}(V_D - V_C e^{ik_u}) + \frac{1}{i\omega L_1}(V_D - V_C e^{-ik_u}) \\
&\quad + \frac{1}{i\omega L_1}(V_D - V_B e^{ik_v}) + \frac{1}{i\omega L_1}(V_D - V_D e^{ik_v}) + i\omega C_1(V_D - V_B e^{-ik_v}) \\
&\quad + \frac{1}{i\omega L_1}(V_D - V_D e^{-ik_v}) + 2i\omega C_1(V_D - 0) + i\omega C_2(V_D - 0).
\end{aligned} \tag{A11}
$$

Equation (A11) can be written in the form of $\mathbf{I} = J\mathbf{V}$. When the driving frequency is set as $\omega = \omega_0$, the circuit Laplacian in momentum space can be written as

$$J = i\sqrt{\frac{C_1}{L_1}}\begin{pmatrix} e^{ik_x} + e^{-ik_x} + e^{ik_z} + e^{ik_y} + e^{-ik_z} - 4 & e^{ik_x} + e^{-ik_z} - e^{ik_x} + e^{-ik_z} & e^{ik_x} + e^{-ik_x} - e^{ik_z} + e^{ik_x} & 0 \\ e^{ik_x} + e^{-ik_x} + e^{ik_z} - e^{-ik_z} & -e^{ik_x} - e^{-ik_x} - e^{ik_z} - e^{-ik_z} - e^{ik_x} - e^{-ik_x} + 4 - \sqrt{2} & 0 & e^{ik_x} - e^{-ik_x} - e^{ik_z} + e^{ik_x} \\ e^{ik_x} + e^{-ik_x} - e^{ik_z} - e^{-ik_x} & 0 & -e^{ik_x} - e^{-ik_x} - e^{ik_z} - e^{-ik_x} - e^{ik_x} - e^{-ik_z} + 4 & -e^{ik_x} - e^{-ik_x} + e^{ik_z} - e^{ik_x} \\ 0 & e^{ik_x} + e^{-ik_x} + e^{ik_z} - e^{-ik_x} & -e^{ik_x} - e^{-ik_x} - e^{ik_z} + e^{ik_x} & e^{ik_x} + e^{-ik_x} + e^{ik_z} + e^{-ik_x} + e^{ik_x} + e^{-ik_x} - 4 + \sqrt{2} \end{pmatrix}.$$

(A12)

It is exactly the same as Eq. (1) in the main text when $a = 1/\sqrt{8}$, $\eta = 1/2$, and $m = 1/2 + 1/(2\sqrt{2})$.

## APPENDIX B: SECOND CHERN NUMBER OF YANG MONOPOLE AND LINKED WEYL SURFACES

Here, we show how to calculate the second Chern number of the Yang monopole and the linked Weyl surfaces in the 5D system. The Yang monopole or Weyl surface is enclosed by a four-dimensional (4D) manifold, and the corresponding topological invariants can be obtained by integrating the Berry curvature on this closed 4D manifold.

Similar to Ref. [22] in the main text, we have $(\kappa - a)^2 + \zeta_4^2 + \zeta_5^2 = \varepsilon_3^2$, where $\kappa = \sqrt{\zeta_1^2 + \zeta_2^2 + \zeta_3^2}$. We can see that if the value of $\varepsilon_3$ is very small, this equation represents a small 4D sphere surrounding the Yang monopole having a topology of $S^4$ when $a = 0$. When $0 < a < 1$, it represents a 4D manifold $\nu$ wrapping the Weyl surface near the Fermi energy, having a topology of $S^2 \times S^2$. We define $E^{i\psi}\sin\alpha = (\zeta_4 + i\zeta_5)/\varepsilon_3$, $\cos\alpha = (\kappa - a)/\varepsilon_3$, $E^{i\phi}\sin\theta = (\zeta_1 + i\zeta_2)/\kappa$, $\cos\theta = \zeta_3/\kappa$. The Hamiltonian $H(\mathbf{k})$ in the main text becomes

$$H = \begin{pmatrix} -(\kappa - \varepsilon_3\cos\alpha) + \kappa\cos\theta & \kappa E^{-i\phi}\sin\theta & E^{-i\psi}\sin\theta & 0 \\ E^{i\phi}\kappa\sin\theta & -(\kappa - \varepsilon_3\cos\alpha) - \eta\cos\theta & 0 & E^{-i\psi}\varepsilon_3\sin\alpha \\ E^{i\psi}\varepsilon_3\sin\alpha & 0 & (\kappa - \varepsilon_3\cos\alpha) - \eta\cos\theta & -E^{-i\phi}\kappa\sin\theta \\ 0 & E^{i\psi}\varepsilon_3\sin\alpha & -\kappa E^{i\phi}\sin\theta & (\kappa - \varepsilon_3\cos\alpha) + \kappa\cos\theta \end{pmatrix}. \quad (B1)$$

For the case of the Yang monopole $a = 0$, the band structure is fourfold degenerate on the Yang monopole and twofold degenerate everywhere else. We should define a $U(2)$ Berry connection as $A_i^{mn} = i\langle u_K^m|\partial K_i|u_K^n\rangle$, with $|u_K^m\rangle$ ($m = 3, 4$). From the solution of Eq. (B2), the two degenerate wave functions corresponding to the energy eigenvalue $\varepsilon_3$ are

$$|u_{\mathbf{k}}^3\rangle = \begin{pmatrix} e^{-i(\phi+\psi)}\cos\frac{\alpha}{2}\cos\frac{\theta}{2} \\ e^{-i\psi}\sin\frac{\alpha}{2}\cos\frac{\theta}{2} \\ e^{-i\phi}\sin\frac{\alpha}{2}\cos\frac{\theta}{2} \\ \sin\frac{\alpha}{2}\sin\frac{\theta}{2} \end{pmatrix}, \quad |u_{\mathbf{k}}^4\rangle = \begin{pmatrix} -e^{-i(\phi+\psi)}\sin\frac{\alpha}{2}\sin\frac{\theta}{2} \\ e^{-i\psi}\sin\frac{\alpha}{2}\cos\frac{\theta}{2} \\ -e^{-i\phi}\cos\frac{\alpha}{2}\sin\frac{\theta}{2} \\ \cos\frac{\alpha}{2}\cos\frac{\theta}{2} \end{pmatrix}. \quad (B2)$$

Then the components of the $U(2)$ Berry connection are

$$A_\phi = \begin{pmatrix} \cos^2\frac{\theta}{2} & -\frac{1}{2}\sin\alpha\sin\theta \\ -\frac{1}{2}\sin\alpha\sin\theta & \sin^2\frac{\theta}{2} \end{pmatrix}, \quad A_\alpha = 0,$$

$$A_\theta = \begin{pmatrix} 0 & -\frac{i}{2}\sin\alpha \\ \frac{i}{2}\sin\alpha & 0 \end{pmatrix}, \quad A_\psi = \begin{pmatrix} \cos^2\frac{\alpha}{2} & 0 \\ 0 & \sin^2\frac{\alpha}{2} \end{pmatrix}. \quad (B3)$$

The non-Abelian $U(2)$ Berry curvature can be expressed by $F_{ij} = \partial k_i A_j - \partial k_j A_i - i[A_i, A_j]$, and we get

$$F_{\alpha\psi} = \begin{pmatrix} -\frac{\sin\alpha}{2} & 0 \\ 0 & \frac{\sin\alpha}{2} \end{pmatrix}, \quad F_{\theta\phi} = \begin{pmatrix} -\frac{1}{2}\cos^2\alpha\sin\theta & 0 \\ 0 & -\frac{1}{2}\cos^2\alpha\sin\theta \end{pmatrix},$$

$$F_{\alpha\phi} = \begin{pmatrix} 0 & -\frac{1}{2}\cos\alpha\sin\theta \\ -\frac{1}{2}\cos\alpha\sin\theta & 0 \end{pmatrix}, \quad F_{\theta\psi} = \begin{pmatrix} 0 & \frac{1}{2}\cos\alpha\sin\alpha \\ \frac{1}{2}\cos\alpha\sin\alpha & 0 \end{pmatrix}, \quad (B4)$$

$$F_{\alpha\theta} = \begin{pmatrix} 0 & -\frac{i}{2}\cos\alpha \\ -\frac{i}{2}\cos\alpha & 0 \end{pmatrix}, \quad F_{\phi\psi} = \begin{pmatrix} 0 & -\frac{i}{2}\cos\alpha\sin\alpha\sin\theta \\ \frac{i}{2}\cos\alpha\sin\alpha\sin\theta & 0 \end{pmatrix},$$

with $F_{ij} = -F_{ji}$. The non-Abelian second Chern number of the Yang monopole can be evaluated by

$$C_2^{NA} = \oint_{S^4} \frac{d^4k\,\varepsilon^{ijkl}[\text{tr}(F_{ij}F_{kl}) - (\text{tr}F_{ij})(\text{tr}F_{kl})]}{32\pi^2}$$

$$= \int_0^\pi d\theta \int_0^\pi d\phi \int_0^\pi d\alpha \int_0^{2\pi} d\psi \frac{\varepsilon^{ijkl}[\text{tr}(F_{ij}F_{kl}) - (\text{tr}F_{ij})(\text{tr}F_{kl})]}{32\pi^2}$$

$$= 1. \quad (B5)$$

For the case with linked Weyl surfaces $0 < a < 1$, the band structure is twofold degenerate only on the Weyl surfaces. The wave function corresponding to the third eigenvalue $\varepsilon_3$ is still $|u_k^3\rangle$ as shown in Eq. (B2), and we use it to calculate the Abelian Chern number of one of the spherelike Weyl surfaces. The $U(1)$ Berry connection is expressed as $A_i^m = i\langle u_K^m|\partial K_i|u_K^m\rangle$, with $|u_K^m\rangle$ the $m$th eigenstates. We have $A_\theta = A_\alpha = 0$, $A_\phi = \cos^2(\theta/2)$, and $A_\psi = \cos^2(\alpha/2)$. The Abelian $U(1)$ Berry curvature is expressed as $F_{ij} = \partial_k i A_j - \partial_k j A_i$. We have $F_{\theta\phi} = -\sin\theta/2$, $F_{\alpha\psi} = -\sin\alpha/2$, $F_{\theta\phi} = -F_{\phi\theta}$, and $F_{\alpha\psi} = -F_{\psi\alpha}$, and the other components of the **B**erry curvature are zero. The Abelian second Chern number

$$
\begin{aligned}
C_2^A &= \oint_v d^4k \frac{\varepsilon^{ijkl}F_{ij}F_{kl}}{32\pi^2} \\
&= \int_0^\pi d\theta \int_0^{2\pi} d\phi \int_0^\pi d\alpha \int_0^{2\pi} d\psi \frac{\varepsilon^{ijkl}F_{ij}F_{kl}}{32\pi^2} \\
&= 1.
\end{aligned} \tag{B6}
$$

## APPENDIX C: DETAILED FABRICATION OF THE 5D CIRCUIT

In order to complete the design of the 5D circuit, two innovations have been made. First, to simulate the Hamiltonian of the 5D Weyl semimetal in the circuit, the operational amplifier usually needs to be used to implement the couplings with the complex values [32]. Unfortunately, the operational amplifier is often not accurate enough, especially in a large circuit chip. It can cause errors and ruin our results. In order to avoid the shortcoming of the operational amplifier, we design a lattice model with all couplings of real values. That is to say, only capacitors and inductors are used in our designed circuits. Second, we design one "switchable" circuit Laplacian in the circuit network. By changing the grounding elements, we can observe the 5D Yang monopole, linked Weyl surfaces, and 5D topological phase transitions in one circuit.

For the values of capacitors and inductors, we choose $L_1 = 330\,\mu H$ and $C_1 = 10$ nF. These values are large enough so that they have good resistance to the stray capacitances/inductances. We select $L_2 = \sqrt{2}L_1 \approx 470\,\mu H$ and use the capacitors of 6.8 nF and 270 pF in parallel to be equivalent to $C_2 = C_1/\sqrt{2}$.

In the experimental design, there are only two plaquettes along the $y$ and $v$ directions. When the periodic boundary condition is applied in these two directions, the inductor $L_1$ represents the coupling with one positive value and the capacitor $C_1$ represents the coupling with the opposite value between two nodes. Therefore, it is equivalent to no couplings between these two nodes. Our experiment is designed with this equivalence; see Figs. 6(a) and 6(b).

Based on the character of the translation invariance, our experimental setup is composed of a number of identical printed circuit board (PCB) modules. The whole experiment setup is shown in Fig. 7(a), and $8\times8\times4\times2$ plaquettes are properly arranged on $8\times8\times2$ PCBs. Every PCB contains $4\times2$ plaquettes and connects with each other through a pin header connector, female header connector, and flat cables. For clarity, we omit the periodic boundary condition in some direction which can be easily recovered by connecting the edge of the experiment setup using flat cables according to the actual experimental situation. Figure 7(b) is two PCBs connected in two layers. At the edge of every PCB are the pin and female header connectors as shown in (1) and (2), which can be used to connect with other PCBs in one horizontal plane. The front view of one PCB is shown Fig. 7(c) where the red dashed rectangle represents a plaquette containing four nodes. The back view of one PCB is shown in Fig. 7(d). (3) shows the grounding elements. Several extra bonding pads in the PCB are presented for the change of grounding elements. (4) is the pin header connector to connect the lower layer. Figure 7(e) shows the front view of the lower layer. (5) is the female header connector that is used to connect the upper layer.

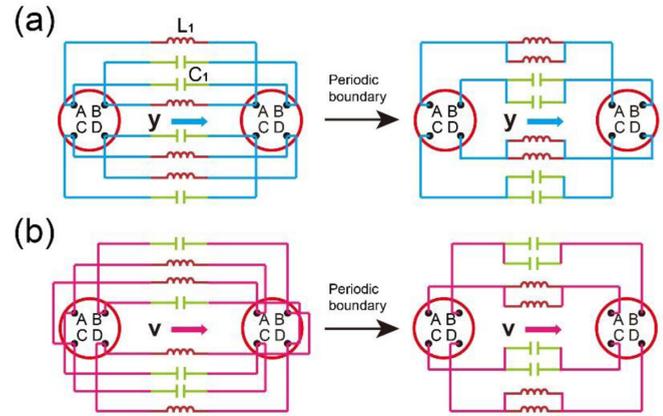

FIG. 6. (a), (b) are the simplification along the $y$ and $v$ directions, respectively, in the experiment under the periodic boundary condition.

## APPENDIX D: RECONSTRUCTION OF THE BAND STRUCTURE IN CIRCUITS

Here, we introduce the method of reconstructing the circuit Laplacian $J$ in momentum space by an elementary voltage measurement and Fourier transform. For the circuit with $N$ nodes, the current flowing into the circuit at the node $i$ is denoted by $I_i$, and the voltage at the node $j$ is denoted by $V_j$. From the Kirchhoff's law, we get $\mathbf{I} = J(\mathbf{r})\mathbf{V}$, where $J(\mathbf{r})$ represents the circuit Laplacian in real space. The method of reconstructing the band structure in momentum space is actually measuring the circuit Laplacian $J(\mathbf{r})$ in real space; then after the Fourier transform, the circuit Laplacian $J(\mathbf{k})$ in momentum space is obtained, and we can calculate the admittance spectrum. So the goal is how to obtain the circuit Laplacian $J(\mathbf{r})$ by measuring the circuit.

First, we do some transformation to $\mathbf{I} = J\mathbf{V}$. We left-multiply both sides of this equation by $G$ which is the inverse matrix of $J$, and we get $G\mathbf{I} = \mathbf{V}$. Here $G$ is called the impedance matrix. Then we excite only one node of the circuit such as at the node $j$, and then measure the input current $I_j$ and the voltages $V_i$ of all $N$ nodes. One column of the impedance matrix can be obtained by $G_{ij} = V_i/I_j$. After exciting the remaining nodes in the circuit and repeating the measurement of the input current and voltages for $N$ nodes, we get the whole impedance matrix $G$. Finally, we take the inverse of $G$ and get $J$.

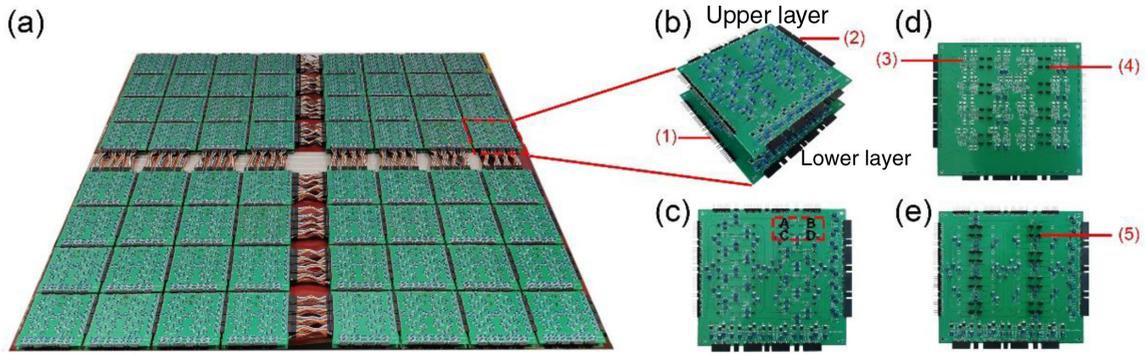

FIG. 7. (a) The whole experiment setup. (b) Two layers of the PCBs. (c) The front view of the upper layer. (d) The back view of the upper layer. (e) The front view of the lower layer.

## APPENDIX E: DISCUSSION ABOUT THE EFFECT OF SAMPLE SIZE

In our experiment, we use the size of the circuit with $8 \times 2 \times 8 \times 4 \times 2$ plaquettes. In this section, we show the feasibil-ity of choosing only two plaquettes in the $y$ and $v$ directions. Corresponding to our real experiment in the main text, here we use the circuit simulation software LTSPICE for virtual experiments. In the following, $N_i$ ($i = x, y, z, u, v$) is used to represent the number of plaquettes in the $i$ directions. We only change $N_y$ and $N_v$ but keep the plaquette number of the other directions as $N_x = 8$, $N_z = 8$, and $N_u = 4$.

We have explained the correspondence between the edge state and the impedance in the main text. Therefore, to illustrate the effect of $N_y$ and $N_v$ on the edge effect, we measure the impedance on the edge of the $v$ direction in the Yang monopole system under different $N_y$ and $N_v$. In Fig. 8(a), we can see that near the resonant frequency, the impedance peak becomes more obvious with the increase of $N_v$. This is because we choose the $v$ direction as the open boundary so that $N_v$ greatly affects the edge effect. Anyway, even though the impedance peak is not perfect when $N_y = N_v = 2$, it can still be observed.

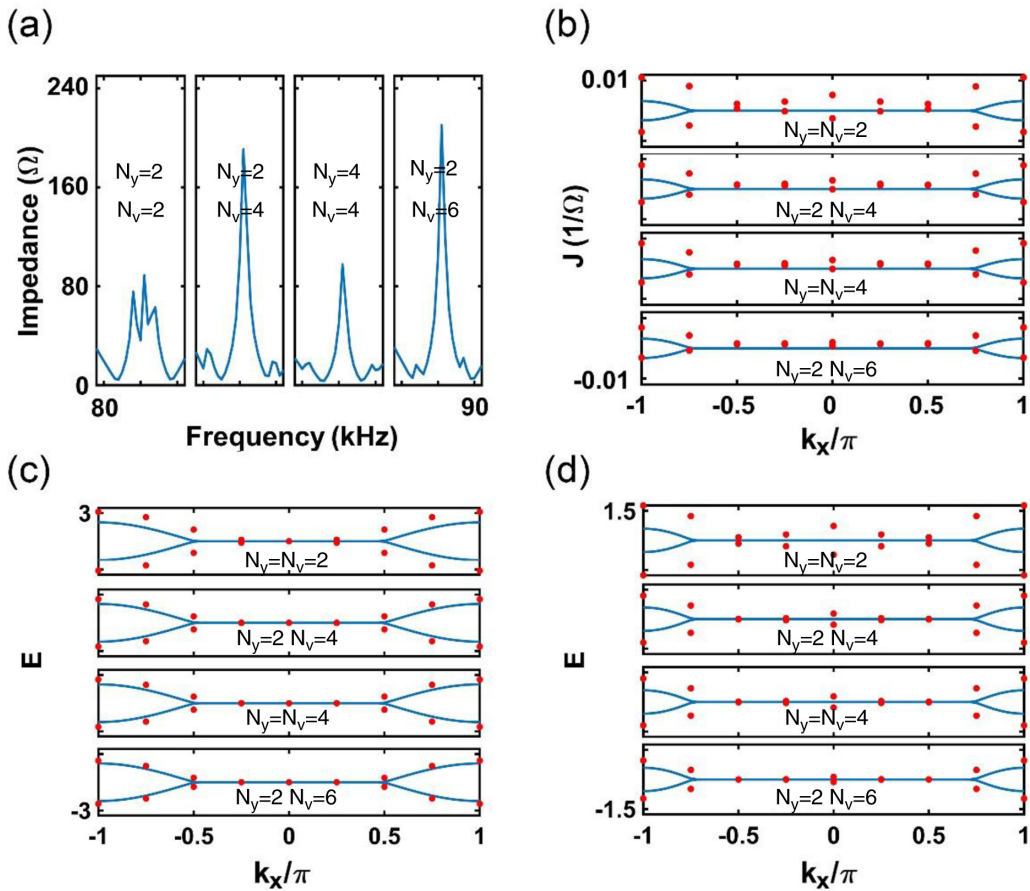

FIG. 8. (a) Impedances measured under different $N_y$ and $N_v$ in the simulated Yang monopole system. (b) Weyl arcs observed under different $N_y$ and $N_v$ in the simulated linked Weyl surface system. (c), (d) are the Weyl arcs observed under different $N_y$ and $N_v$ in the tight-binding Yang monopole and linked Weyl surface system, respectively.

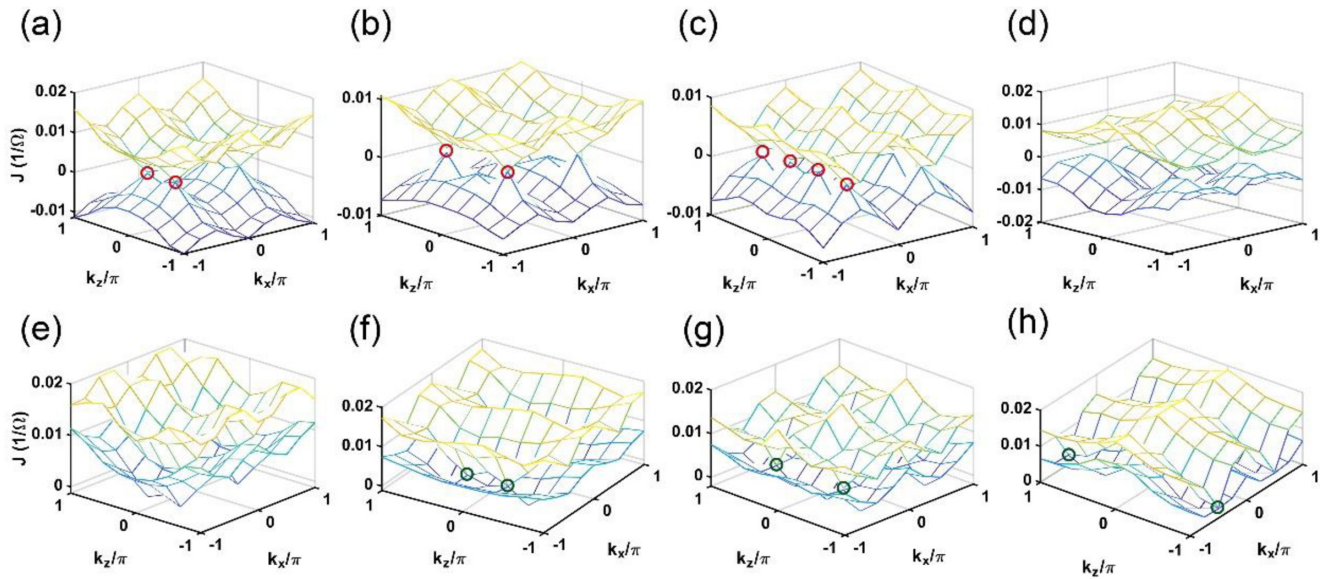

FIG. 9. (a)–(d) show the admittance spectra of bands $\varepsilon_2$ and $\varepsilon_3$ we measured in the real experiment corresponding to the phases of Figs. 4(b)–4(e) in the main text, while (e)–(h) are the admittance spectra between bands $\varepsilon_3$ and $\varepsilon_4$ corresponding to the phases of Figs. 4(b)–4(e) in the main text.

The effect of $N_y$ and $N_v$ on Weyl arcs is investigated in the linked Weyl surfaces system; the result is similar to the edge effect case. Four panels in Fig. 8(b) show the Weyl arcs we can observe under different $N_y$ and $N_v$. We can see that the more plaquettes in the $v$ direction, the more obvious the Weyl arc, and increasing $N_y$ does not affect the Weyl arc. Even in the case with $N_y = N_v = 2$, we can see the shape of the Weyl arc. Therefore, we conclude that two plaquettes along the $y$ and $v$ directions are enough to observe the key points in the band structure in our experiment. We also show the theory calculation results about the Weyl arc observed in the tight-binding model of the Yang monopole and linked Weyl surfaces in Figs. 3(c) and 3(d), respectively. The theory calculation results are in agreement with the experimental and simulation results.

## APPENDIX F: DETAILED ADMITTANCE SPECTRA IN THE EXPERIMENT OF PHASE TRANSITION

In the section, we show the measured admittance spectra in the subspace of $k_x$ and $k_z$. We still use $\varepsilon_i$ to represent the $i$th band. Figs. 9(a)–9(d) show the admittance spectra of bands $\varepsilon_2$ and $\varepsilon_3$ being measured in the experiment, which corresponds to the phases of Figs. 4(b)–4(e) in the main text, While Figs. 9(e)–9(h) show the admittance spectra between bands $\varepsilon_3$ and $\varepsilon_4$ corresponding to the phases of Figs. 4(b)–4(e) in the main text. The band crossing points correspond to the red and green dots in Fig. 4 that are marked by red and green circles here.